\documentclass[final,5p,times,twocolumn]{elsarticle}
\usepackage{lineno,hyperref,amsmath,mathtools,amsthm,amssymb,amsfonts,ragged2e,color,subfig}
\journal{``Contributions to Plasma Physics''}
\begin{document}
\begin{frontmatter}
\title{Obliquely propagating ion-acoustic shock waves in degenerate quantum plasma}
\author{M.K. Islam$^{1,*}$, S. Biswas$^{1,**}$, N.A. Chowdhury$^{2,***}$, A. Mannan$^{1,\dag}$, M. Salahuddin$^{1,\ddag}$, and A.A. Mamun$^{1,\S}$}
\address{$^1$ Department of Physics, Jahangirnagar University, Savar, Dhaka-1342, Bangladesh\\
$^{2}$Plasma Physics Division, Atomic Energy Centre, Dhaka-1000, Bangladesh\\
e-mail: $^{*}$islam.stu2018@juniv.edu, $^{**}$shawonbiswas440@gmail.com, $^{***}$nurealam1743phy@gmail.com,\\
$^{\dag}$abdulmannan@juniv.edu, $^{\ddag}$su\_2960@juniv.edu, $^{\S}$mamun\_phys@juniv.edu}
\begin{abstract}
A theoretical investigation has been carried out on the propagation of nonlinear ion-acoustic shock waves (IASHWs) in a
collsionless magnetized degenerate quantum plasma system composed of  inertial non-relativistic positively charged light
and heavy ions, inertialess ultra-relativistically degenerate electrons and positrons. The reductive perturbation method
has been employed to drive the Burgers' equation. It has been observed that under consideration, our plasma model supports only
positive potential shock structure. It is also found that the amplitude and steepness of the IASHWs have been significantly modified
by the variation of ion kinematic viscosity, oblique angle, number density, and
charge state of the plasma species.  The results of our present investigation will be helpful for understanding
the propagation of IASHWs in white dwarfs and neutron stars.
\end{abstract}
\begin{keyword}
Shock waves; Degenerate quantum plasma; White dwarfs; Neutron stars.
\end{keyword}
\end{frontmatter}
\section{Introduction}
\label{2sec:Introduction}
The existence of light (viz., ${}^{1}_{1}{\mbox{H}}$ \cite{Chandrasekhar1931a,Chandrasekhar1934,Fletcher2006,Killian2006}, ${}^{4}_{2}{{\mbox{He}}}$ \cite{Chandrasekhar1931a,Horn1991,Fowler1994,Koester1990,Koester2002}, ${}^{12}_{6}{\mbox{C}}$ \cite{Koester1990,Koester2002}, and ${}^{16}_{8}{\mbox{O}}$ \cite{Koester1990,Koester2002,Chandrasekhar1931b}, etc.) and heavy (viz., ${}^{56}_{26}{\mbox{Fe}}$ \cite{Vanderburg2015,Chamel2008}, ${}^{85}_{37}{\mbox{Rb}}$ \cite{Chamel2008,Witze2014}, and ${}^{96}_{42}{\mbox{Mo}}$ \cite{Chamel2008,Witze2014}, etc.) ions in astrophysical compact objects (viz., white dwarfs and neutron stars) has received a substantial attention to investigate ion-acoustic (IA) waves (IAWs) in degenerate quantum plasma (DQP). The particle number density in the white dwarfs (i.e., $6\times 10^{29}\mbox{cm}^{-3}$) and neutron stars (i.e., $6\times 10^{36}\mbox{cm}^{-3}$) is extremely high \cite{Chandrasekhar1931a,Chandrasekhar1931b,Chamel2008,Chandrasekhar1964}. This extreme number density of these degenerate particles dictates to follow the  Heisenberg's uncertainty principle, and according to the uncertainty principle when the position of these
particles is confined then the momentum of these particles tends to very large. This excessive
momentum leads to generate extreme outward degenerate pressure which is counter-balanced by the inward gravitational compression.
The dynamics of these particles is mathematically modeled under two categories, namely, non-relativistic and ultra-relativistic limits by Chandrasekhar \cite{Chandrasekhar1931a}. The pressure of non-relativistic light and heavy ions is expressed as $P_j=K_jN_j^{\alpha}$; where $\alpha=5/3$, $K_j=3\pi \hbar^2/5 m_j$,  $j=l$ for light ion, $j=h$ for heavy ion, and $\hbar$ is the Planck constant \cite{Shukla2011,El-Taibany2012a,Sultana2018}.
The degenerate pressure of ultra-relativistic electrons and positrons can be expresses as  $P_s=K_sN_s^{\gamma_s}$;
where $s=e$ for electron, $s=p$ for positron, $\gamma_s=4/3$, $K_s=3\hbar c/4$, and $c$ is the speed of
light \cite{Shukla2011,El-Taibany2012a,Sultana2018}.

The presence of positrons in white dwarfs and neutron stars has been extensively discussed in the Refs. \cite{Zhang2005,Sturrock1971,Harding1998,Harding2006}. Sultana and Schlickeiser \cite{Sultana2018} investigated IA solitary waves in DQP containing degenerate electrons, light ions, and inertial mobile non-degenerate heavy ions. Gill \textit{et al.} \cite{Gill2010} investigated the IA shock waves (IASHWs) in relativistic DQP composed of electrons, positrons and ions, and found that the height of the potential is maximum for the lower positron density. Ata-ur-Rahman \textit{et al.} \cite{Ata-Ur-Rahman2013b} considered an unmagnetized DQP containing inertial ions, and inertialess electrons and positrons, and demonstrated that the amplitude of the IAWs decreases with the increase of positron number density. Hossen \textit{et al.} \cite{Hossen2017a} studied IASHWs in a four-component plasma system having inertialess electrons and positrons, and inertial heavy and light ions, and reported that the amplitude of the shock profile decreases with positron number density. Mamun and Shukla \cite{Mamun2010} investigated electrostatic solitary waves propagating in ultra-relativistic plasma medium consisting of degenerate electrons and cold mobile ions, and reported that the wave amplitude increases with the increase of ion number density.

The strong magnetic field (i.e., about 1 Mega Gauss) in white dwarfs was predicted by Blackett \cite{Blackett1947} and observed by Zeeman spectroscopy \cite{Liebert1977,Euchner2002}. El-Taibany \textit{et al.} \cite{El-Taibany2012b} analyzed solitary waves in a magnetized degenerate electron-positron plasma, and found that the wave amplitude increases with the oblique angle which is the angle between the external magnetic field and the direction of wave propagation. Shaukat \cite{Shaukat2017} investigated IA solitary waves in the presence of magnetic field, and observed that the solitary wave amplitude increases with increasing oblique angle.

The energy dissipation of the shock wave, which is governed by the Burgers' equation \cite{Burgers1948}, may arise due to the kinematic viscosity of the medium. Hafez \textit{et al.} \cite{Hafez2017} studied IASHWs in weakly relativistic plasma containing electrons, positrons, and ions, and noticed that the steepness  of the IASHWs decreases with the increase in the value of viscosity of plasma species but the amplitude of the IASHWs does not change. Abdelwahed \textit{et al.} \cite{Abdelwahed2016} analyzed IASHWs in a pair-ion plasma, and also found that the shock steepness decreases with increasing ion viscosity.

Recently, Saini \textit{et al.} \cite{Saini2020} investigated heavy nucleus acoustic periodic waves in DQP. Haider \cite{Haider2016}
examined the shock profiles in the presence of degenerate inertial ions, and inertialess electrons and positrons. Atteya \textit{et al.} \cite{Atteya2017} studied IASHWs in a DQP which contains ion fluids, degenerate electrons, and stationary heavy ions. To the best of authors' knowledge, still no one investigated IASHWs in a magnetized DQP having inertialess ultra-relativistically degenerate electrons and positrons, and inertial positively charged non-relativistic light and heavy ions. In this manuscript, we will derive the Burgers' equation, and will also use associated solution to examine the basic features of IASHW in DQP.

The manuscript is organized in the preceding way: The governing equations are described
in section \ref{2sec:Governing Equations}. The Burgers' equation and associated shock solution
are presented in section \ref{2sec:Derivation of Burgers equation}. The results and discussion
are presented in section \ref{2sec:Results and Discussions}. A brief conclusion is presented
in section \ref{2sec:Conclusion}.
\section{Governing Equations}
\label{2sec:Governing Equations}
We consider a magnetized DQP system consisting of inertial positively charged light ion (mass $m_l$; charge $eZ_l$; number density $N_l$),
positively charged heavy ion (mass $m_h$; charge $eZ_h$, number density $N_h$), inertialess electron (mass $m_e$; charge $-e$, number density $N_e$), and positron (mass $m_p$; charge $e$; number density $N_p$);
where $Z_l$ ($Z_h$) is the charge state of the light (heavy) ion. An uniform external magnetic field $\mathbf{B}$ is existing in the direction of $z$-axis ($\mathbf{B}={B_0}\hat{z}$ and $\hat{z}$ is the unit vector). The propagation of IAWs in DQP system is governed by the following equations:
\begin{eqnarray}
&&\hspace*{-1.3cm}\frac{\partial N_{h}}{\partial T}+\tilde\nabla \cdot(N_{h}{U}_{h})=0,
\label{2eq:1}\\
&&\hspace*{-1.3cm}\frac{\partial U_{h}}{\partial T}+({U}_{h}\cdot\tilde\nabla){U}_{h}=-\frac{Z_{h}e}{m_{h}}\nabla\tilde{\Phi} +\frac{Z_{h}eB_0}{m_{h}}({U}_{h}\times\hat{z})
\nonumber\\
&&\hspace*{1.7cm}-\frac{1}{m_{h}N_{h}}\tilde{\nabla}P_{h}+\tilde\eta_h\tilde\nabla^2{U}_h,
\label{2eq:2}\\
&&\hspace*{-1.3cm}\frac{\partial N_{l}}{\partial T}+\tilde\nabla \cdot(N_{l}{U}_{l})=0,
\label{2eq:3}\\
&&\hspace*{-1.3cm}\frac{\partial {U}_{l}}{\partial T}+({U}_{l}\cdot\tilde\nabla){U}_{l}=-\frac{Z_{l}e}{m_{l}}\tilde\nabla\tilde{\Phi} +\frac{Z_{l}eB_0}{m_{l}}({U}_{l}\times\hat{z})
\nonumber\\
&&\hspace*{1.52cm}-\frac{1}{m_{l}N_{l}}\tilde{\nabla}P_{l}+\tilde\eta_l\tilde\nabla^2{U}_l,
\label{2eq:4}\\
&&\hspace*{-1.3cm}\tilde\nabla^2\tilde\Phi=4\pi e(N_e-N_p-Z_lN_l-Z_hN_h),
\label{2eq:5}\
\end{eqnarray}
where ${U}_l$ (${U}_h$) is the fluid speed of light (heavy) ion; $\tilde\Phi$ is the electrostatic wave potential; $P_l$ ($P_h$)
is the pressure for light (heavy) ion;  $\tilde\eta_l=\mu/m_lN_l$ ($\tilde\eta_h=\mu/m_hN_h$) is the kinematic viscosity for light (heavy) ion.
The degenerate pressure equations for electrons and positrons can be expressed, respectively, as
\begin{eqnarray}
&&\hspace*{-1.3cm}\tilde\nabla\tilde\Phi-\frac{1}{eN_e}\tilde\nabla P_e=0,
\label{2eq:6}\\
&&\hspace*{-1.3cm}\tilde\nabla\tilde\Phi+\frac{1}{eN_p}\tilde\nabla P_p=0.
\label{2eq:7}\
\end{eqnarray}
Now, we have introduced the normalizing parameters: $n_h\rightarrow N_h/n_{h0}$; $n_l\rightarrow N_l/n_{l0}$; $n_e\rightarrow N_e/n_{e0}$; $n_p\rightarrow N_p/n_{p0}$; $u_h\rightarrow U_h/C_h$; $u_l\rightarrow U_l/C_h$; $\phi\rightarrow e\tilde\Phi/m_ec^2$; $t\rightarrow\omega_{ph}T$; $\nabla\rightarrow \lambda_{Dh}^{-1}\tilde\nabla$; $\eta\rightarrow \tilde\eta/\omega_{ph}\lambda_{Dh}^2$ \big[where IAWs speed $C_h=(Z_hm_ec^2/m_h)^{1/2}$; plasma frequency $\omega_{ph}=(4\pi Z_h^2e^2n_{h0}/m_h)^{1/2}$; the Debye length $\lambda_{Dh}=(m_ec^2/4\pi Z_he^2n_{h0})^{1/2}$, and for simplicity we have considered $\eta=\eta_l=\eta_h$]. At equilibrium, the charge neutrality condition can be written as $n_{e0}=n_{p0}+Z_ln_{l0}+Z_hn_{h0}$. By using these normalizing parameters, Eqs. \eqref{2eq:1}-\eqref{2eq:5} can be expressed in the normalized form
\begin{eqnarray}
&&\hspace*{-1.3cm}\frac{\partial n_{h}}{\partial t}+\nabla \cdot(n_{h}u_{h})=0,
\label{2eq:8}\\
&&\hspace*{-1.3cm}\frac{\partial}{\partial t}(u_{h})+(u_{h}\cdot\nabla)u_{h}=-\nabla\phi +\Omega_{ch}(u_{h}\times\hat{z})
\nonumber\\
&&\hspace*{1.8cm}-\frac{\mu_1K_1}{n_{h}}\nabla n_{h}^\alpha+\eta\nabla^2{u_{h}},
\label{2eq:9}\\
&&\hspace*{-1.3cm}\frac{\partial n_{l}}{\partial t}+\nabla \cdot(n_{l}u_{l})=0,
\label{2eq:10}\\
&&\hspace*{-1.3cm}\frac{\partial}{\partial t}(u_{l})+(u_{l}\cdot\nabla)u_{l}=-\mu_2\nabla\phi +\mu_2\Omega_{ch}(u_{l}\times\hat{z})
\nonumber\\
&&\hspace*{1.6cm}-\frac{\mu_3K_2}{n_{l}}\nabla n_{l}^\alpha+\eta\nabla^2{u_{l}},
\label{2eq:11}\\
&&\hspace*{-1.3cm}\nabla^2\phi=(1+\mu_4+\mu_5)n_e-\mu_4n_{p}-\mu_5n_l-n_h,
\label{2eq:12}\
\end{eqnarray}
where $\Omega_{ch}=\omega_{ch}/\omega_{ph}$, $\mu_1=1/Z_h$, $\mu_2=Z_lm_h/Z_hm_l$, $\mu_3=m_h/Z_hm_e$, $\mu_4=n_{p0}/Z_{h}n_{h0}$, $\mu_5=Z_ln_{l0}/Z_{h}n_{h0}$, $K_1={n_{h0}}^{\alpha-1}K_h/m_ec^2$, and $K_2={n_{l0}}^{\alpha-1}K_l/m_ec^2$.  Now, by normalizing and integrating Eqs. \eqref{2eq:6} and \eqref{2eq:7}, the number densities of the inertialess electrons and positrons can be obtained in terms of electrostatic potential $\phi$, respectively, as
\begin{eqnarray}
&&\hspace{-1.3cm}n_e=\bigg[1+\frac{\gamma_e-1}{\gamma_eK_3}\phi\bigg]^{\frac{1}{\gamma_e-1}},
\label{2eq:13}\\
&&\hspace{-1.3cm}n_p=\bigg[1-\frac{\gamma_p-1}{\gamma_pK_4}\phi\bigg]^{\frac{1}{\gamma_p-1}},
\label{2eq:14}\
\end{eqnarray}
where $K_3={n_{e0}}^{\gamma_e-1}K_e/m_ec^2$ and $K_4={n_{p0}}^{\gamma_p-1}K_p/m_ec^2$. By expanding the right hand side of Eqs. \eqref{2eq:13} and \eqref{2eq:14} up to second order in $\phi$, and substituting in Eq. \eqref{2eq:12}, we get
\begin{eqnarray}
&&\hspace*{-1.3cm}\nabla^2\phi+\mu_5n_l+n_h=1+\mu_5+\beta_1\phi+\beta_2\phi^2+\cdot\cdot\cdot,
\label{2eq:15}\
\end{eqnarray}
where
\begin{eqnarray}
&&\hspace*{-1.3cm}\beta_1=\frac{[\gamma_pK_4(1+\mu_4+\mu_5)+\mu_4\gamma_eK_3]}{\gamma_e\gamma_pK_3K_4},
\nonumber\\
&&\hspace*{-1.3cm}\beta_2=\frac{[(\gamma_pK_4)^2(1+\mu_4+\mu_5)(2-\gamma_e)-\mu_4(\gamma_eK_3)^2(2-\gamma_p)]}{2(\gamma_e\gamma_pK_3K_4)^2}.
\nonumber\
\end{eqnarray}
\section{Derivation of the Burgers' Equation}
\label{2sec:Derivation of Burgers equation}
To study IASHWs, we derive Burgers' equation by employing the reductive perturbation method (RPM) \cite{C1,C2},
and the stretched coordinates for independent variables can be written as \cite{Mamun1999,Hossen2017b,Washimi1966}
\begin{eqnarray}
&&\hspace*{-1.3cm} \xi=\epsilon(l_xx+l_yy+l_zz-v_pt),
\label{2eq:16}\\
&&\hspace*{-1.3cm} \tau=\epsilon^2t,
\label{2eq:17}\
\end{eqnarray}
where $v_p$ is the phase speed and $\epsilon$ is a smallness parameter measuring the
weakness of the dissipation (0$<$$\epsilon$$<$1). The $l_x$, $l_y$, and $l_z$ are the
directional cosines of $\boldsymbol{k}$ (wave vector) along $x$, $y$, and $z$-axes,
respectively (i.e., $l_x^2+l_y^2+l_z^2=1$). Then, the dependent variables can be expressed
in power series of $\epsilon$ as \cite{Hossen2017b}
\begin{eqnarray}
&&\hspace*{-1.3cm} n_h=1+\epsilon n_h^{(1)}+\epsilon^2n_h^{(2)}+\cdot\cdot\cdot,
\label{2eq:18}\\
&&\hspace*{-1.3cm} n_l=1+\epsilon n_l^{(1)}+\epsilon^2n_l^{(2)}+\cdot\cdot\cdot,
\label{2eq:19}\\
&&\hspace*{-1.3cm} u_{hx,y}=\epsilon^2 u_{hx,y}^{(1)}+\epsilon^3 u_{hx,y}^{(2)}+\cdot\cdot\cdot,
\label{2eq:20}\\
&&\hspace*{-1.3cm} u_{lx,y}=\epsilon^2 u_{lx,y}^{(1)}+\epsilon^3 u_{lx,y}^{(2)}+\cdot\cdot\cdot,
\label{2eq:21}\\
&&\hspace*{-1.3cm} u_{hz}=\epsilon u_{hz}^{(1)}+\epsilon^2 u_{hz}^{(2)}+\cdot\cdot\cdot,
\label{2eq:22}\\
&&\hspace*{-1.3cm} u_{lz}=\epsilon u_{lz}^{(1)}+\epsilon^2 u_{lz}^{(2)}+\cdot\cdot\cdot,
\label{2eq:23}\\
&&\hspace*{-1.3cm}\phi=\epsilon \phi^{(1)}+\epsilon^2 \phi^{(2)}+\cdot\cdot\cdot.
\label{2eq:24}\
\end{eqnarray}
Now, by substituting Eqs. \eqref{2eq:16}-\eqref{2eq:24} into Eqs. \eqref{2eq:8}-\eqref{2eq:11} and \eqref{2eq:15},
and collecting the terms containing $\epsilon$, the first-order equations reduce to
\begin{eqnarray}
&&\hspace*{-1.3cm}n_h^{(1)} = \frac{ l_z^2}{\big(v_p^2 - \alpha\mu_1 l_z^2 K_1\big)}\phi^{(1)},
\label{2eq:25}\\
&&\hspace*{-1.3cm}u_{hz}^{(1)} = \frac{v_p l_z}{\big(v_p^2 - \alpha\mu_1 l_z^2 K_1\big)}\phi^{(1)},
\label{2eq:26}\\
&&\hspace*{-1.3cm}n_l^{(1)} = \frac{\mu_2 l_z^2}{\big(v_p^2 - \alpha\mu_3 l_z^2 K_2\big)}\phi^{(1)},
\label{2eq:27}\\
&&\hspace*{-1.3cm}u_{lz}^{(1)} = \frac{\mu_2 v_p l_z}{\big(v_p^2 - \alpha\mu_3 l_z^2 K_2\big)}\phi^{(1)}.
\label{2eq:28}\
\end{eqnarray}
Now, the phase speed of IASHWs can be written as
\begin{eqnarray}
&&\hspace*{-1.3cm}v_p=v_{p+}= l_z \sqrt{\frac{a + \sqrt{a^2 -4 \beta_1 b}}{2 \beta_1}},
\label{2eq:29}\\
&&\hspace*{-1.3cm}v_p = v_{p-}= l_z \sqrt{\frac{a - \sqrt{a^2 -4 \beta_1 b}}{2 \beta_1}},
\label{2eq:30}\
\end{eqnarray}
where $a=\alpha\mu_1\beta_1 K_1+\alpha\mu_3\beta_1K_2 +\mu_2\mu_5+1$ and
$b=\alpha^2\mu_1\mu_3\beta_1K_1K_2-\alpha\mu_1\mu_2\mu_5K_1-\alpha\mu_3K_2$.
The $x$ and $y$-components of the first-order momentum equations can be manifested as
\begin{eqnarray}
&&\hspace*{-1.3cm}u_{hx}^{(1)}=-\frac{l_y v_p^2}{\Omega_{ch}(v_p^2 - \alpha\mu_1 l_z^2 K_1)} \frac{\partial\phi^{(1)}}{\partial\xi},
\label{2eq:31}\\
&&\hspace*{-1.3cm}u_{hy}^{(1)} =  \frac{l_x v_p^2}{\Omega_{ch}(v_p^2 - \alpha\mu_1 l_z^2 K_1)} \frac{\partial\phi^{(1)}}{\partial\xi},
\label{2eq:32}\\
&&\hspace*{-1.3cm}u_{lx}^{(1)}=-\frac{l_y v_p^2}{\Omega_{ch}(v_p^2 - \alpha\mu_3 l_z^2 K_2)} \frac{\partial\phi^{(1)}}{\partial\xi},
\label{2eq:33}\\
&&\hspace*{-1.3cm}u_{ly}^{(1)} =  \frac{l_x v_p^2}{\Omega_{ch}(v_p^2 - \alpha\mu_3 l_z^2 K_2)} \frac{\partial\phi^{(1)}}{\partial\xi}.
\label{2eq:34}\
\end{eqnarray}
Now, by taking the next higher-order terms, the equation of
continuity, momentum equation, and Poisson’s equation can be written as
\begin{eqnarray}
&&\hspace*{-1.3cm} \frac{ \partial n_h^{(1)}}{\partial \tau} - v_p  \frac{ \partial n_h^{(2)}}{\partial \xi}
+ l_x \frac{ \partial u_{hx}^{(1)}}{\partial \xi} + l_y \frac{ \partial u_{hy}^{(1)}}{\partial \xi}+ l_z \frac{ \partial u_{hz}^{(2)}}{\partial \xi}
\nonumber\\
&&\hspace*{0cm}+ l_z \frac{\partial}{\partial \xi} \Big(n_h^{(1)}  u_{hz}^{(1)} \Big) = 0,
\label{2eq:35}\\
&&\hspace*{-1.3cm} \frac{ \partial u_{hz}^{(1)}}{\partial \tau} - v_p \frac{ \partial u_{hz}^{(2)}}{\partial \xi} + l_z  u_{hz}^{(1)} \frac{\partial u_{hz}^{(1)}}{\partial \xi} + \mu_1 l_z \frac{\partial \phi^{(2)}}{\partial \xi}
\nonumber\\
&&\hspace*{-1cm}+ \alpha\mu_1 l_z K_1 \Biggr[ \frac{\partial n_h^{(2)}}{\partial \xi}
+ \frac{(\alpha-2)}{2}  \frac{\partial n_{h}^{(1)^2}}{\partial \xi} \Biggr] - \eta \frac{\partial^2 u_{hz}^{(1)}}{\partial \xi^2} = 0,
\label{2eq:36}\\
&&\hspace*{-1.3cm} \frac{ \partial n_l^{(1)}}{\partial \tau} - v_p  \frac{ \partial n_l^{(2)}}{\partial \xi}
+ l_x \frac{ \partial u_{lx}^{(1)}}{\partial \xi} + l_y \frac{ \partial u_{ly}^{(1)}}{\partial \xi} + l_z \frac{ \partial u_{lz}^{(2)}}{\partial \xi}
\nonumber\\
&&\hspace*{0cm}+ l_z \frac{\partial}{\partial \xi} \Big(n_l^{(1)}  u_{lz}^{(1)} \Big) = 0,
\label{2eq:37}\\
&&\hspace*{-1.3cm} \frac{ \partial u_{lz}^{(1)}}{\partial \tau} - v_p \frac{ \partial u_{lz}^{(2)}}{\partial \xi} +  l_z u_{lz}^{(1)}\frac{\partial u_{lz}^{(1)}}{\partial \xi} +\mu_2l_z \frac{\partial \phi^{(2)}}{\partial \xi}
\nonumber\\
&&\hspace*{-1cm}+ \alpha\mu_3l_z K_2 \Biggr[ \frac{\partial n_l^{(2)}}{\partial \xi}
+\frac{(\alpha-2)}{2}  \frac{\partial n_{l}^{(1)^2}}{\partial \xi} \Biggr] - \eta \frac{\partial^2 u_{lz}^{(1)}}{\partial \xi^2} = 0,
\label{2eq:38}\\
&&\hspace*{-1.3cm}\mu_5n_l^{(2)}+n_h^{(2)}=\beta_1\phi^{(2)}+\beta_2\phi^{(1)^2}.
\label{2eq:39}\
\end{eqnarray}
Finally, the next higher-order terms of Eqs. \eqref{2eq:8}-\eqref{2eq:11} and \eqref{2eq:15},
with the help of Eqs. \eqref{2eq:25}-\eqref{2eq:39}, can provide the Burgers' equation
\begin{eqnarray}
&&\hspace*{-1.3cm}\frac{\partial \Phi}{\partial \tau}+A\Phi\frac{\partial \Phi}{\partial \xi}=B\frac{\partial^2 \Phi}{\partial \xi^2},
\label{2eq:40}\
\end{eqnarray}
where $\Phi=\phi^{(1)}$ for simplicity. In Eq. \eqref{2eq:32}, the nonlinear
coefficient $A$ and dissipative coefficient $B$ are given
\begin{eqnarray}
&&\hspace*{-1.3cm}A=A'(A''-2\beta_2),
\label{2eq:41}\\
&&\hspace*{-1.3cm}B=\frac{\eta}{2},
\label{2eq:42}\
\end{eqnarray}
where
\begin{eqnarray}
&&\hspace*{-1.4cm}A'= \frac{v_p^4+\alpha^2\mu_1\mu_3 l_z^4K_1 K_2-\alpha l_z^2 v_p^2\big(\mu_1K_1+\mu_3K_2\big)}{2v_p l_z^2\big[(v_p^2-\alpha\mu_3l_z^2K_2)^2+\mu_2\mu_5(v_p^2-\alpha\mu_1l_z^2K_1)^2\big]},
\nonumber\\
&&\hspace*{-1.4cm}A''=\frac{l_z^4\big[\{3v_p^2+\mu_1l_z^2K_1\alpha(\alpha-2)\}+\mu_2^2\mu_5\{3v_p^2+\mu_3l_z^2K_2\alpha(\alpha-2)\}\big]}{\big\{(v_p^2-\alpha\mu_1l_z^2K_1)(v_p^2-\alpha\mu_3l_z^2K_2)\big\}^3},
\nonumber\
\end{eqnarray}
Now, we look for stationary shock wave solution of this Burgers' equation by considering
$\zeta=\xi-U_0\tau$ (where $\zeta$ is a new space variable and $U_0$ is the speed of the ion fluid).
These allow us to write the stationary shock wave solution as
\begin{eqnarray}
&&\hspace*{-1.3cm}\Phi=\Phi_0\bigg[1-\tanh\bigg(\frac{\zeta}{\Delta}\bigg)\bigg],
\label{2eq:43}\
\end{eqnarray}
where the amplitude $\Phi_0$ and width $\Delta$ are, respectively, given by
\begin{eqnarray}
&&\hspace*{-1.3cm}\Phi_0=\frac{U_0}{A},~~~\mbox{and}~~~\Delta=\frac{2B}{U_0}.
\label{2eq:44}\
\end{eqnarray}
It is clear from Eqs. \eqref{2eq:43} and \eqref{2eq:44} that the IASHWs exist, which
are formed due to the balance between nonlinearity and dissipation, because $B>0$ and the
IASHWs with $\Phi>0$ ($\Phi<0$) exist if $A>0$ ($A<0$) because $U_0>0$.
\begin{figure}[t]
\centering
\includegraphics[width=80mm]{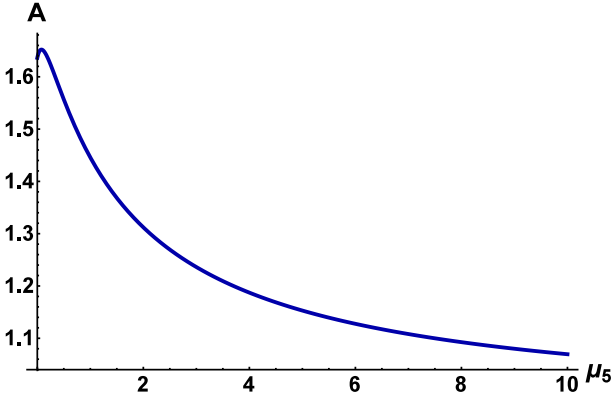}
\caption{Plot of the nonlinear coefficient $A$ vs $\mu_5$ when $\alpha=5/3$, $\gamma_e=4/3$, $\gamma_p=4/3$, $\delta=20^\circ$, $\eta=0.3$,
$Z_h=37$, $Z_l=6$, $n_{h0}=7\times10^{29}\mbox{cm}^{-3}$, $n_{l0}=5\times 10^{30} \mbox{cm}^{-3}$, $n_{p0}=10^{31}\mbox{cm}^{-3}$, and $v_p=v_{p+}$.}
\label{2Fig:1}
\end{figure}
\begin{figure}[t]
\centering
\includegraphics[width=80mm]{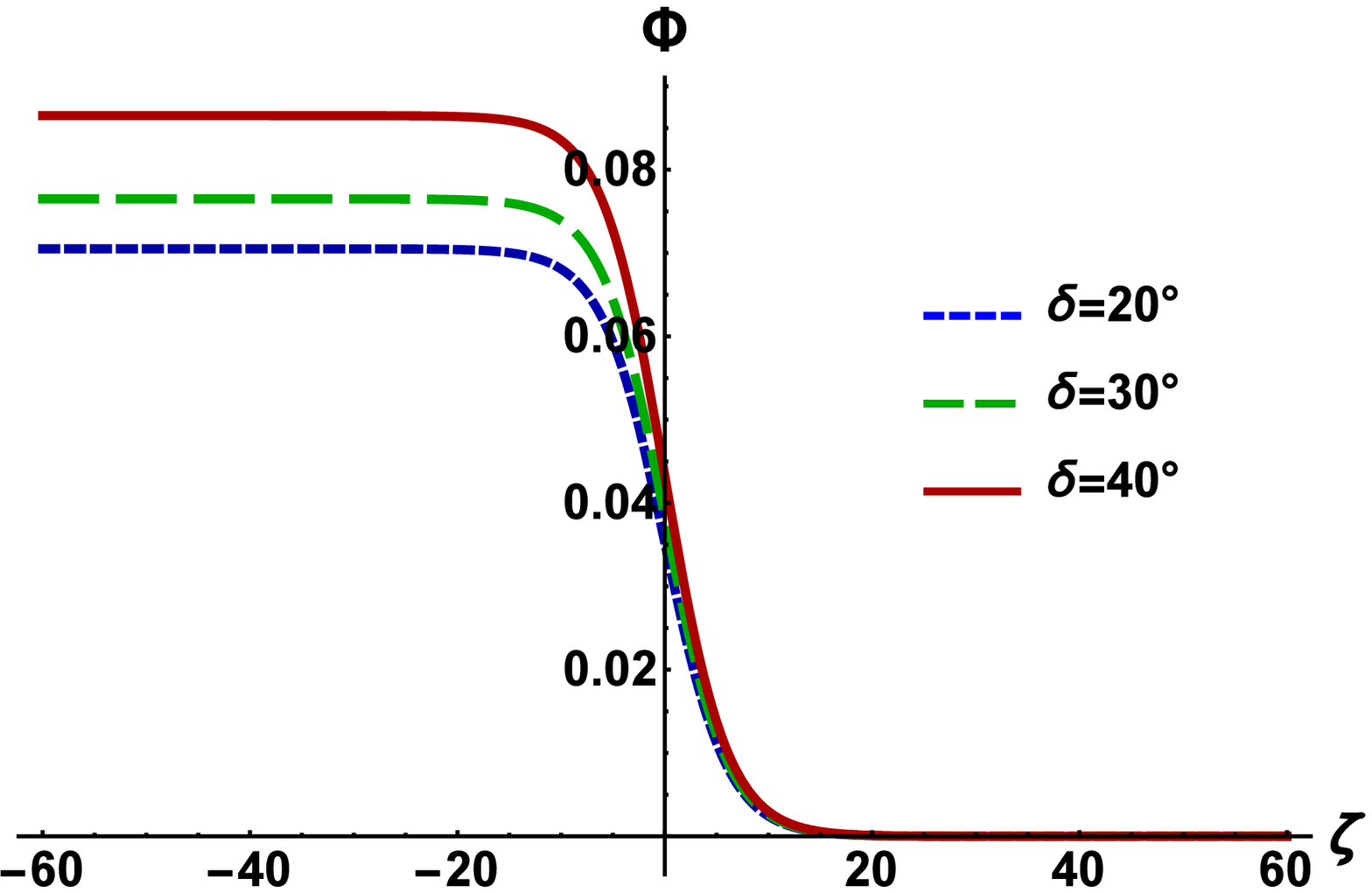}
\caption{Plot of $\Phi$ vs $\zeta$ for different values of $\delta$ when $\alpha=5/3$; $\gamma_e=4/3$, $\gamma_p=4/3$,
$\eta=0.3$, $Z_h=37$, $Z_l=6$, $n_{h0}=7\times10^{29}\mbox{cm}^{-3}$, $n_{l0}=5\times 10^{30} \mbox{cm}^{-3}$,
$n_{p0}=10^{31}\mbox{cm}^{-3}$, $U_0=0.05$, and $v_p=v_{p+}$.}
\label{2Fig:2}
\end{figure}
\begin{figure}[t]
\centering
\includegraphics[width=80mm]{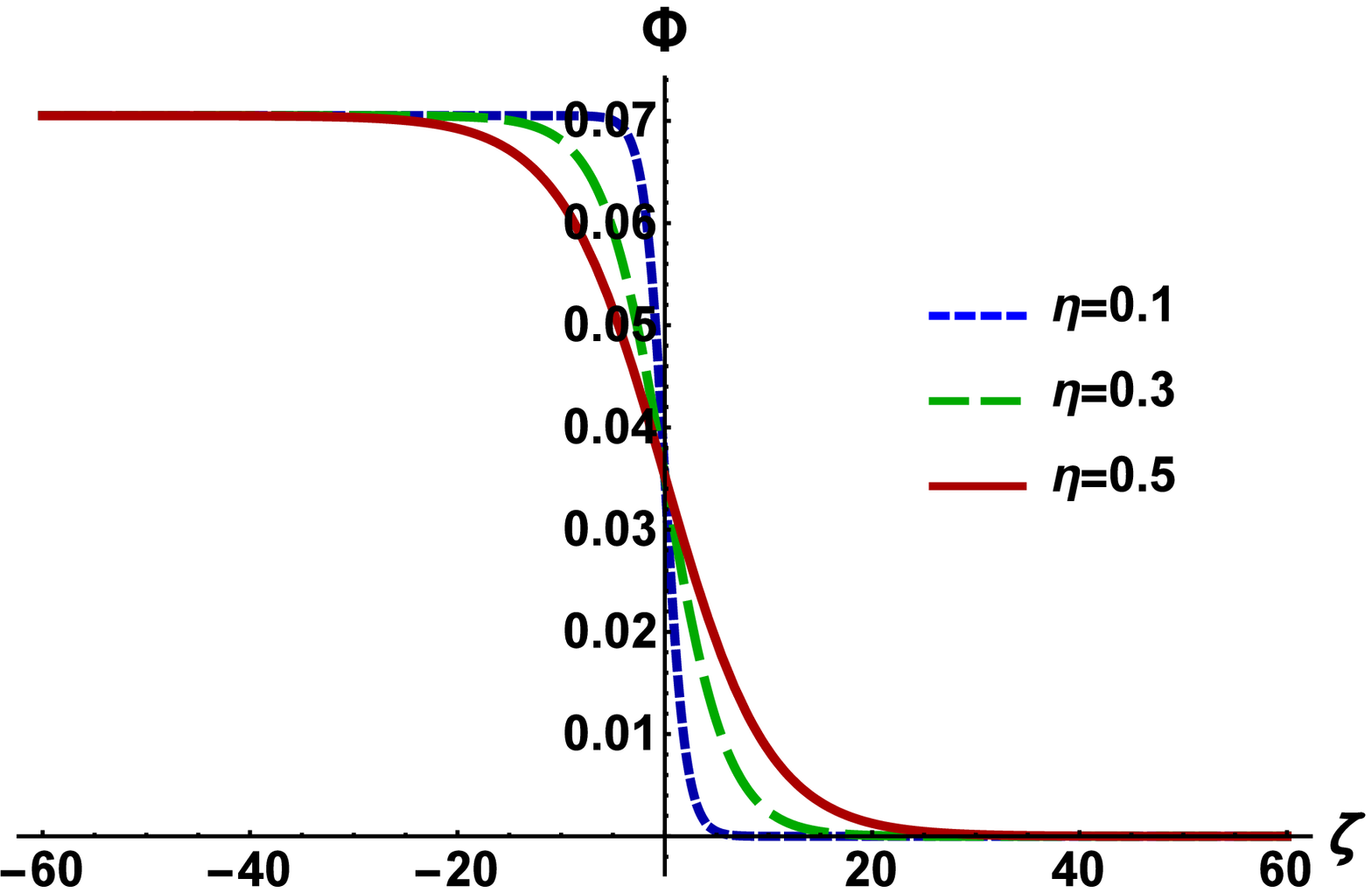}
\caption{Plot of $\Phi$ vs $\zeta$ for different values of $\eta$ when $\alpha=5/3$; $\gamma_e=4/3$, $\gamma_p=4/3$,
$\delta=20^\circ$, $Z_h=37$, $Z_l=6$, $n_{h0}=7\times10^{29}\mbox{cm}^{-3}$, $n_{l0}=5\times 10^{30} \mbox{cm}^{-3}$,
$n_{p0}=10^{31}\mbox{cm}^{-3}$, $U_0=0.05$, and $v_p=v_{p+}$.}
\label{2Fig:3}
\end{figure}
\begin{figure}[t]
\centering
\includegraphics[width=80mm]{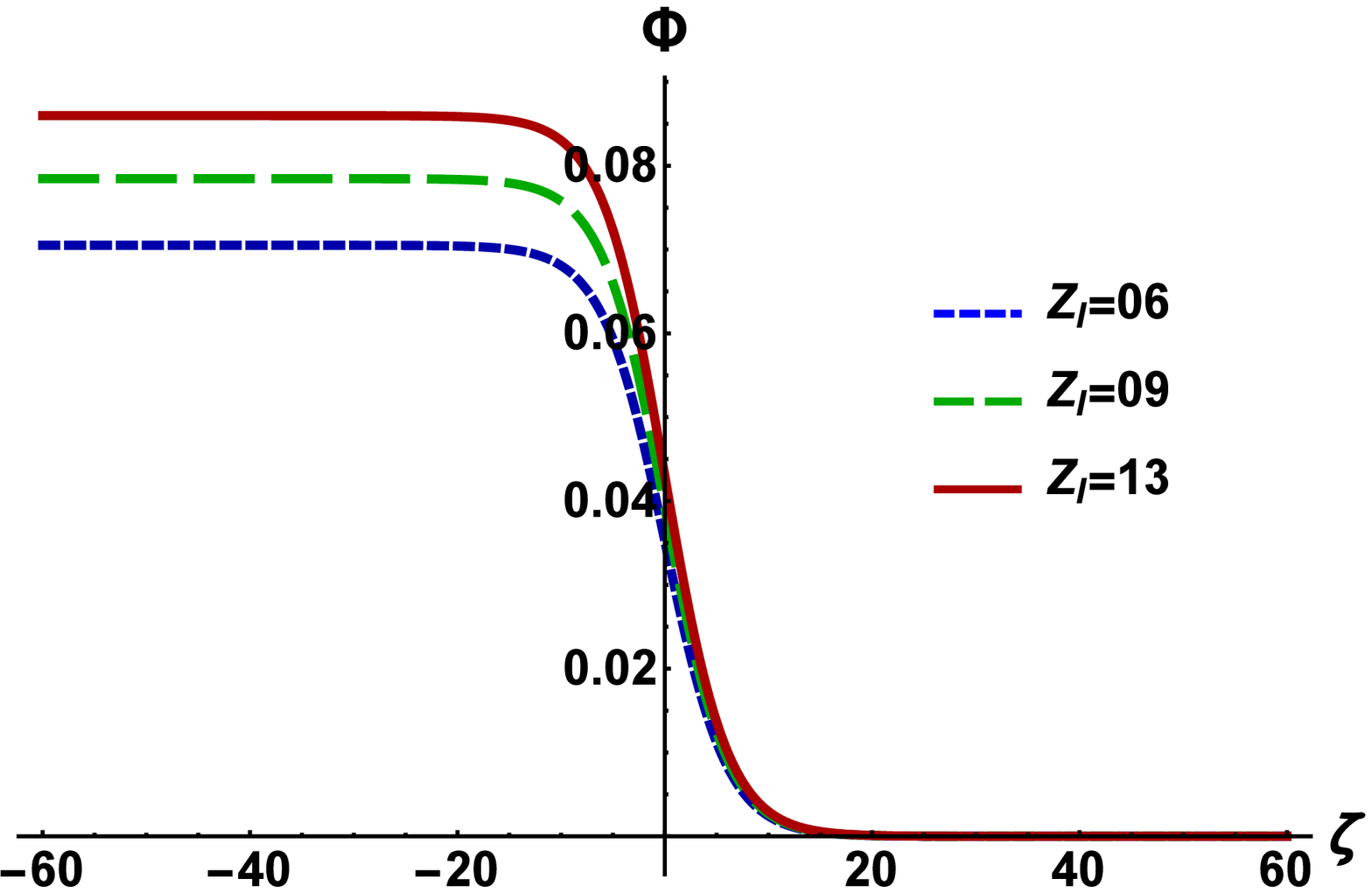}
\caption{Plot of $\Phi$ vs $\zeta$  for different values of $Z_l$ when $\alpha=5/3$; $\gamma_e=4/3$, $\gamma_p=4/3$,
$\delta=20^\circ$, $\eta=0.3$, $Z_h=37$, $n_{h0}=7\times10^{29}\mbox{cm}^{-3}$, $n_{l0}=5\times 10^{30} \mbox{cm}^{-3}$,
$n_{p0}=10^{31}\mbox{cm}^{-3}$, $U_0=0.05$, and $v_p=v_{p+}$.}
\label{2Fig:4}
\end{figure}
\begin{figure}[t]
\centering
\includegraphics[width=80mm]{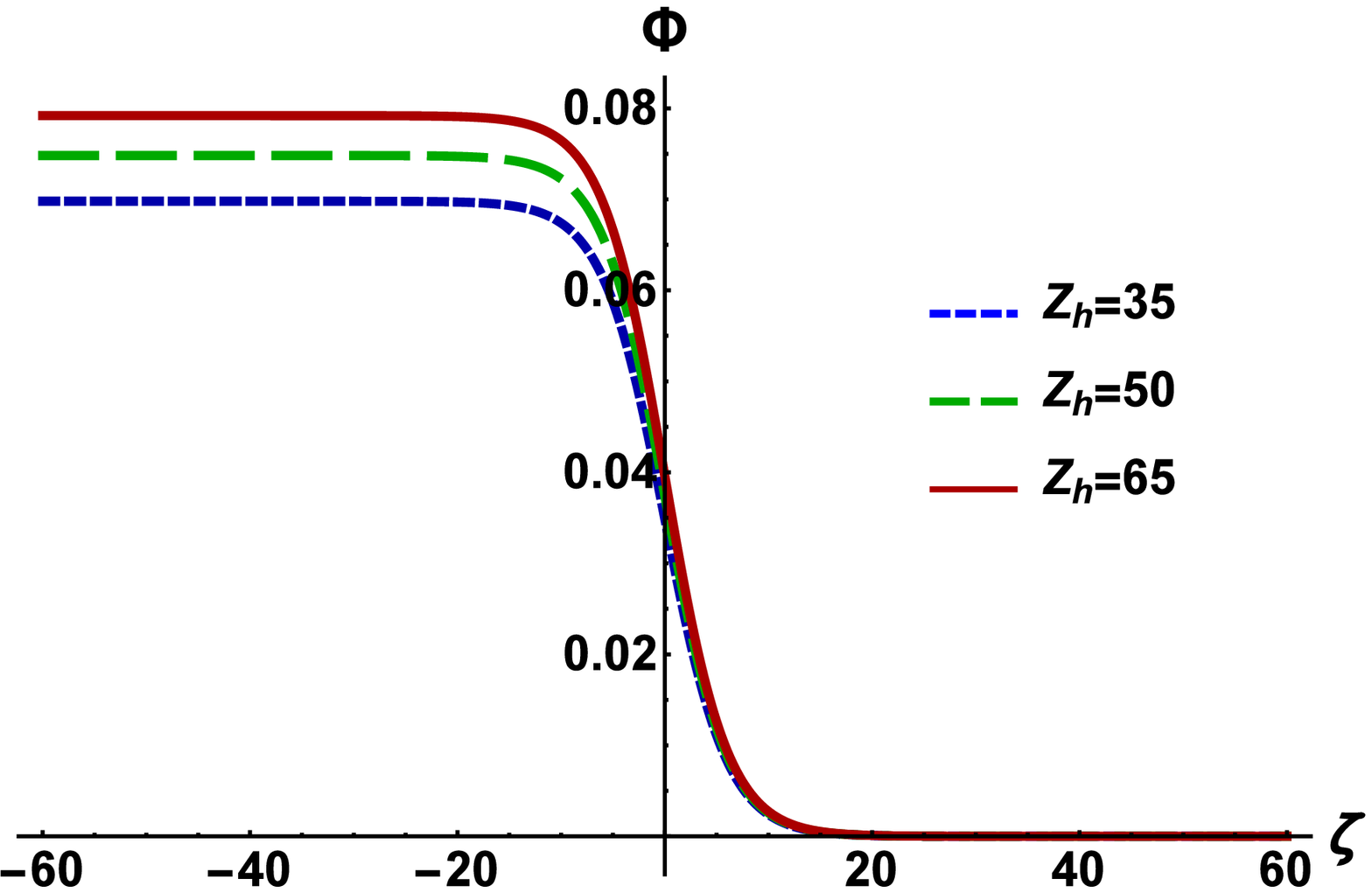}
\caption{Plot of $\Phi$ vs $\zeta$ for different values of $Z_h$ when $\alpha=5/3$; $\gamma_e=4/3$, $\gamma_p=4/3$,
$\delta=20^\circ$, $\eta=0.3$, $Z_l=6$, $n_{h0}=7\times10^{29}\mbox{cm}^{-3}$, $n_{l0}=5\times 10^{30} \mbox{cm}^{-3}$,
$n_{p0}=10^{31}\mbox{cm}^{-3}$, $U_0=0.05$, and $v_p=v_{p+}$.}
\label{2Fig:5}
\end{figure}
\begin{figure}[t]
\centering
\includegraphics[width=80mm]{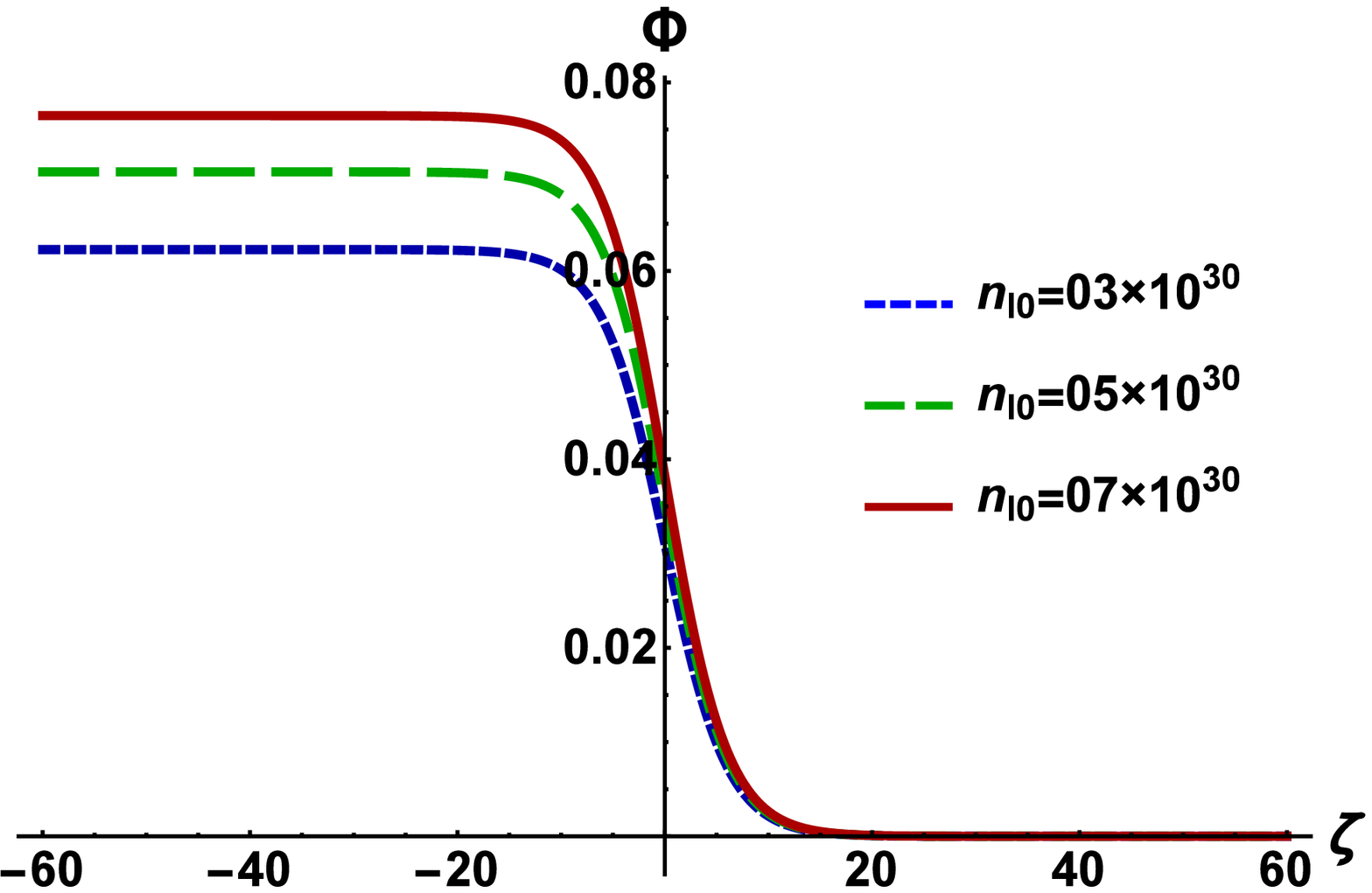}
\caption{Plot of $\Phi$ vs $\zeta$ for different values $n_{l0}$ when $\alpha=5/3$; $\gamma_e=4/3$, $\gamma_p=4/3$,
$\delta=20^\circ$, $\eta=0.3$, $Z_h=37$, $Z_l=6$, $n_{h0}=7\times10^{29}\mbox{cm}^{-3}$,
$n_{p0}=10^{31}\mbox{cm}^{-3}$, $U_0=0.05$, and $v_p=v_{p+}$.}
\label{2Fig:6}
\end{figure}
\begin{figure}[t]
\centering
\includegraphics[width=80mm]{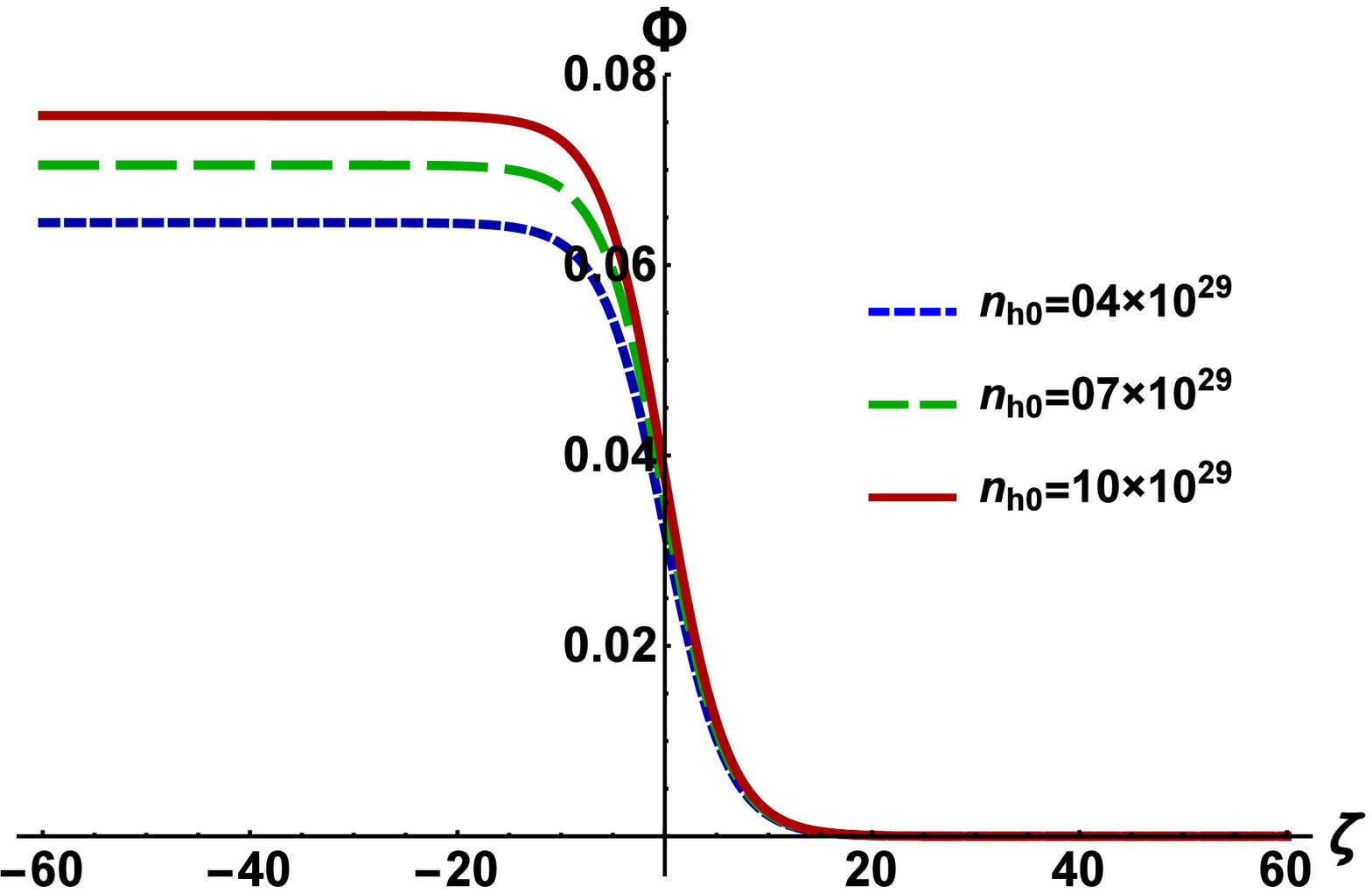}
\caption{Plot of $\Phi$ vs $\zeta$  for different values of $n_{h0}$ when $\alpha=5/3$; $\gamma_e=4/3$, $\gamma_p=4/3$,
$\delta=20^\circ$, $\eta=0.3$, $Z_h=37$, $Z_l=6$, $n_{l0}=5\times 10^{30} \mbox{cm}^{-3}$,
$n_{p0}=10^{31}\mbox{cm}^{-3}$, $U_0=0.05$, and $v_p=v_{p+}$.}
\label{2Fig:7}
\end{figure}
\begin{figure}[t]
\centering
\includegraphics[width=80mm]{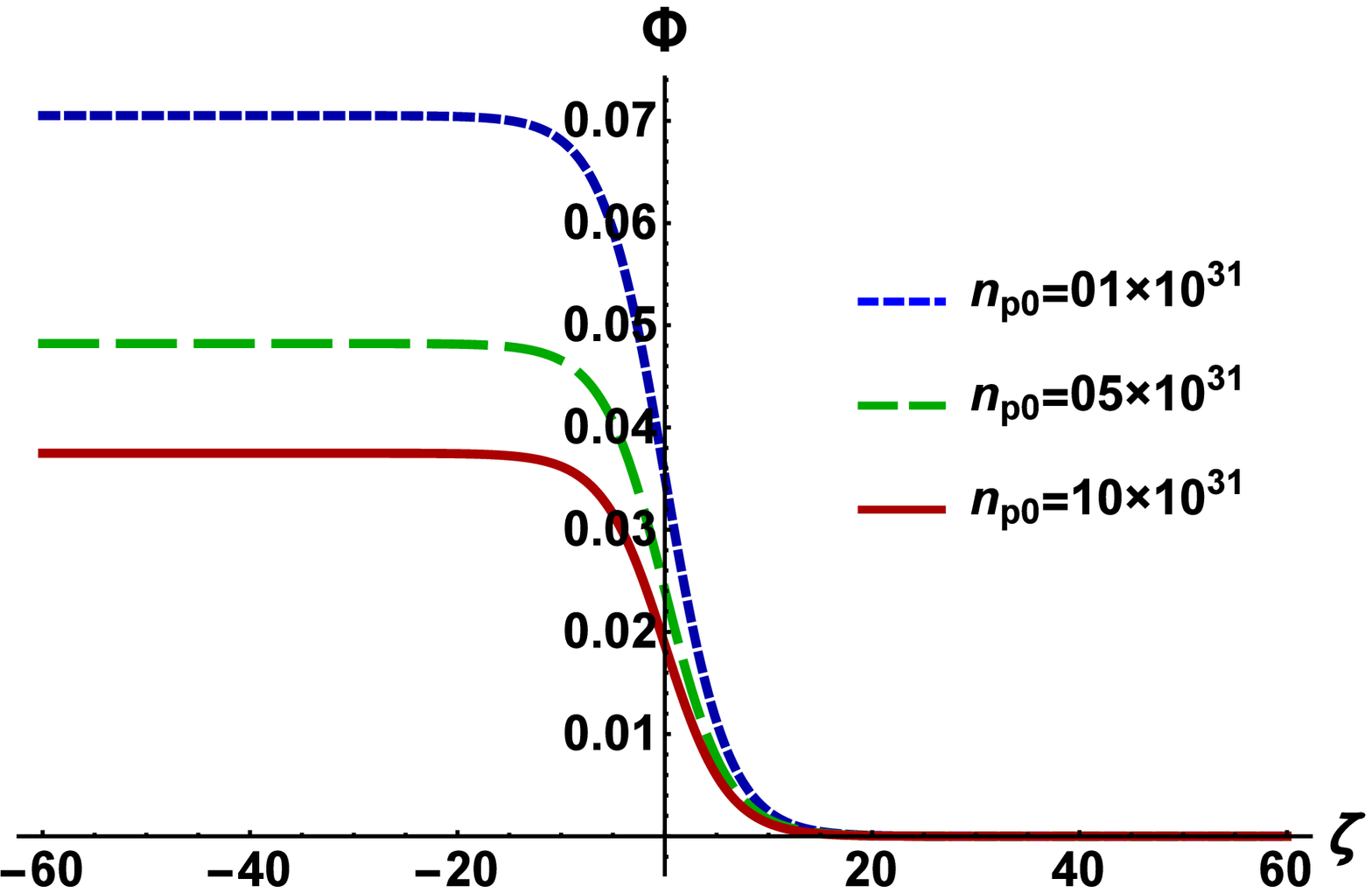}
\caption{Plot of $\Phi$ vs $\zeta$ for different values of $n_{p0}$ when $\alpha=5/3$; $\gamma_e=4/3$, $\gamma_p=4/3$,
$\delta=20^\circ$, $\eta=0.3$, $Z_h=37$, $Z_l=6$, $n_{h0}=7\times10^{29}\mbox{cm}^{-3}$, $n_{l0}=5\times 10^{30} \mbox{cm}^{-3}$,
$U_0=0.05$, and $v_p=v_{p+}$.}
\label{2Fig:8}
\end{figure}
\section{Results and Discussion}
\label{2sec:Results and Discussions}
Our investigation is reasonably valid for degenerate cold plasma systems (viz., white dwarfs and neutron stars)
in which light (viz., ${}^{1}_{1}{\mbox{H}}$ \cite{Chandrasekhar1931a,Chandrasekhar1934,Fletcher2006,Killian2006}, ${}^{4}_{2}{{\mbox{He}}}$ \cite{Chandrasekhar1931a,Horn1991,Fowler1994,Koester1990,Koester2002}, ${}^{12}_{6}{\mbox{C}}$ \cite{Koester1990,Koester2002}, and ${}^{16}_{8}{\mbox{O}}$ \cite{Koester1990,Koester2002,Chandrasekhar1931b}, etc.) and heavy (viz., ${}^{56}_{26}{\mbox{Fe}}$ \cite{Vanderburg2015,Chamel2008}, ${}^{85}_{37}{\mbox{Rb}}$ \cite{Chamel2008,Witze2014}, and ${}^{96}_{42}{\mbox{Mo}}$ \cite{Chamel2008,Witze2014}, etc.) ions can exist. For our numerical analysis,
we have considered the range of the plasma parameters as $Z_l=6\thicksim13$, $Z_h=35\thicksim65$, $n_{l0}=3\times 10^{30} \mbox{cm}^{-3}\thicksim7\times 10^{30} \mbox{cm}^{-3}$, $n_{h0}=4\times 10^{29} \mbox{cm}^{-3}\thicksim10\times 10^{29} \mbox{cm}^{-3}$, and $n_{p0}=1\times 10^{31} \mbox{cm}^{-3}\thicksim10\times 10^{31} \mbox{cm}^{-3}$.

It is obvious from Eq. \eqref{2eq:40} that the wave potential becomes infinite when nonlinear
coefficient is equal to zero (i.e., $A=0$), and in that case, the validity of reductive perturbation method breaks down.
The positive (i.e., $\Phi>0$) and negative (i.e., $\Phi<0$) electrostatic potentials can exist corresponding to the values of $A>0$ and $A<0$, respectively.
It is clear from Fig. \ref{2Fig:1} that our plasma model supports only positive potential shock structure (i.e., $\Phi>0$) associated with $A>0$ under
consideration of non-relativistic light and heavy ions (i.e., $\alpha=5/3$), and ultra-relativistic degenerate electrons and
positrons (i.e., $\gamma_e=\gamma_p=4/3$).

The oblique angle $\delta$ indicates the angle between the direction of the propagation of IASHW and the
direction of existing external magnetic field which is parallel with $z$-axes. The variation
of electrostatic shock potential structure (i.e., $\Phi>0$) associated with $A>0$ with $\delta$ under consideration of non-relativistic
light and heavy ions (i.e., $\alpha=5/3$) and ultra-relativistic degenerate electrons and
positrons (i.e., $\gamma_e=\gamma_p=4/3$) can be observed in Fig. \ref{2Fig:2}. It is obvious from this figure
that the electrostatic shock potential (i.e., $\Phi>0$) associated with $A>0$ increases with the increase of $\delta$.
Physically, the interaction between the electrostatic shock potential associated with $A>0$ and the external magnetic field
is clearly increased with the increase in $\delta$.

Figure \ref{2Fig:3} describes the effects of the kinematic viscosity of the positively charged non-relativistic
light and heavy ions (via $\eta$) on the electrostatic shock potential structure (i.e., $\Phi>0$) associated with $A>0$.
We have noticed that the amplitude of the electrostatic shock profile (i.e., $\Phi>0$) associated with $A>0$ is independent to the variation
of the ion kinematic viscosity but the steepness of the electrostatic shock profile (i.e., $\Phi>0$)
associated with $A>0$ is rigorously dependent to the variation
of ion kinematic viscosity. The steepness of the electrostatic shock potential structure (i.e., $\Phi>0$) associated with $A>0$
decreases with $\eta$, and this result agrees with the result of Refs. \cite{Hafez2017,Abdelwahed2016}.

The amplitude of the positive electrostatic shock structure (i.e., $\Phi>0$) associated with $A>0$ is so much sensitive
to the change of the charge state of non-relativistic light and heavy ions. Figures \ref{2Fig:4} and \ref{2Fig:5} show
that the amplitude of the positive shock structure (i.e., $\Phi>0$) associated with $A>0$ increases with the charge
state of non-relativistic light and heavy ions. Similarly, the increasing number of non-relativistic light and heavy ions
enhances the positive electrostatic shock structure (i.e., $\Phi>0$) associated with $A>0$. It can easily demonstrate from
Figs. \ref{2Fig:6} and \ref{2Fig:7} that as we increase $n_{l0}$ and $n_{h0}$, the amplitude of the positive electrostatic
shock structure (i.e., $\Phi>0$) associated with $A>0$ increases. Physically, the charge state and number density of non-relativistic
light and heavy ions can control the dynamics of the DQP system in the similar way.

We have studied the characteristics of IASHWs for different values of positron number density (via $n_{p0}$)
under consideration of non-relativistic light and heavy ions (i.e., $\alpha=5/3$) and ultra-relativistic degenerate
electrons and positrons (i.e., $\gamma_e=\gamma_p=4/3$) in Fig. \ref{2Fig:8}. It can be highlighted from this
figure that the amplitude of electrostatic shock structure (i.e., $\Phi>0$) associated with $A>0$
decreases with positron number density, and this finding is analogous to the result of Refs. \cite{Gill2010,Ata-Ur-Rahman2013b} .
\section{Conclusion}
\label{2sec:Conclusion}
We have studied the basic characteristics of IASHWs in an extremely dense DQP containing non-relativistic
light and heavy ions, and inertialess ultra-relativistic degenerate electrons and positrons in the presence of external
magnetic field. The RPM \cite{C3} has been utilized to derive Burgers' equation. The results
that have been found from our present investigation can be summarized as follows:
\begin{itemize}
  \item  Our plasma model supports only positive potential shock structure (i.e., $\Phi>0$) associated with $A>0$ under
        consideration of non-relativistic light and heavy ions (i.e., $\alpha=5/3$), and ultra-relativistic degenerate electrons and
        positrons (i.e., $\gamma_e=\gamma_p=4/3$).
  \item The electrostatic shock potential (i.e., $\Phi>0$) associated with $A>0$ increases with the increase in $\delta$.
  \item The steepness of the positive potential shock structure (i.e., $\Phi>0$) associated with $A>0$ decreases with $\eta$.
  \item The amplitude of electrostatic shock structure (i.e., $\Phi>0$) associated with $A>0$ is to be found to increase with the charge
state and number density of non-relativistic light and heavy ions.
  \item The increasing positron number density decreases the height of the positive shock profile.
\end{itemize}
It may be noted that the self-gravitational effect of the plasma species is important to be
considered in our governing equations but beyond the scope of our present work. Overall,
the outcomes from our present investigation will be helpful to understand the IASHWs in white dwarfs and neutron stars.


\begin{thebibliography}{99}

\bibitem{Chandrasekhar1931a} S. Chandrasekhar, Astrophys. J. \textbf{74}, 81 (1931).

\bibitem{Chandrasekhar1934} S. Chandrasekhar, The Observatory \textbf{57}, 373 (1934).

\bibitem{Fletcher2006} R.S. Fletcher, \textit{et al.}, Phys. Rev. Lett. \textbf{96}, 105003 (2006).

\bibitem{Killian2006} T.C. Killian, Nature (London) \textbf{441}, 297 (2006).

\bibitem{Horn1991} H.M. Van Horn, Science \textbf{252}, 384 (1991).

\bibitem{Fowler1994} R.H. Fowler, J. Astrophys. Astron. \textbf{15}, 105 (1994).

\bibitem{Koester1990} D. Koester and G. Chanmugam, Rep. Prog. Phys. \textbf{53}, 837 (1990).

\bibitem{Koester2002} D. Koester, Astron. Astrophys. Rev. \textbf{11}, 33 (2002).

\bibitem{Chandrasekhar1931b} S. Chandrasekhar, Philos. Mag. \textbf{11}, 592 (1931).

\bibitem{Vanderburg2015} A. Vanderburg, \textit{et al.}, Nature \textbf{526}, 546 (2015).

\bibitem{Chamel2008} N. Chamel and P. Haensel, Living Rev. Relativ. \textbf{11}, 10 (2008).

\bibitem{Witze2014} A. Witze, Nature \textbf{510}, 196 (1990).

\bibitem{Chandrasekhar1964} S. Chandrasekhar and R.F. Tooper, Astrophys. J. \textbf{139}, 1396 (1964).

\bibitem{Shukla2011} P.K. Shukla, \textit{et al.}, Phys. Rev. E \textbf{84}, 026405 (2011).

\bibitem{El-Taibany2012a} W.F. El-Taibany, \textit{et al.},  Adv. space res. \textbf{50}, 101 (2012).

\bibitem{Sultana2018} S. Sultana and R. Schlickeiser, Astrophys. Space Sci. \textbf{363}, 103 (2018).

\bibitem{Zhang2005} B. Zhang and J. Gil, Astrophys. J. \textbf{631}, 143 (2005).

\bibitem{Sturrock1971} P.A. Sturrock, Astrophys. J. \textbf{164} (1971) 529.

\bibitem{Harding1998}  A.K. Harding and A. G. Muslimov, Astrophys. J. \textbf{508}, 328 (1998).

\bibitem{Harding2006} A.K. Harding and D. Lai, Rep. Prog. Phys. \textbf{69}, 2631 (2006).

\bibitem{Gill2010} T.S. Gill, \textit{et al.}, J. Phys. \textbf{208}, 012040 (2010).

\bibitem{Ata-Ur-Rahman2013b} Ata-Ur-Rahman, \textit{et al.}, J. Plasma Phys. \textbf{79}, 817 (2013).

\bibitem{Hossen2017a} M.R. Hossen, \textit{et al.},  Plasma Phys. Rep. \textbf{43}, 1189 (2017).

\bibitem{Mamun2010} A.A. Mamun and P.K. Shukla, Phys. Plasmas \textbf{17}, 104504 (2010).

\bibitem{Blackett1947} P.M.S. Blackett, Nature \textbf{159}, 658 (1947).

\bibitem{Liebert1977} J. Liebert, \textit{et al.},  Astrophys. J. \textbf{214}, 457 (1977).

\bibitem{Euchner2002} F. Euchner, \textit{et al.}, Astron. Astrophys. \textbf{390}, 633 (2002).

\bibitem{El-Taibany2012b} W.F. El-Taibany, \textit{et al.}, Adv. Space Res. \textbf{50}, 101 (2012).

\bibitem{Shaukat2017} M.I. Shaukat, Phys. Plasmas \textbf{24}, 102301 (2017).

\bibitem{Burgers1948} J.M. Burgers, Elsevier \textbf{1}, 171 (1948).

\bibitem{Hafez2017} M.G. Hafez, \textit{et al.}, Plasma Phys. Rep. \textbf{43}, 499 (2017).

\bibitem{Abdelwahed2016} H.G. Abdelwahed, \textit{et al.}, J. Exp. Theor. Phys. \textbf{122}, 1111 (2016).

\bibitem{Saini2020} N.S. Saini, \textit{et al.},  Wave Random Complex. DOI: 10.1080/17455030.2020.1798561.

\bibitem{Haider2016} M. M. Haider, Z. Naturforsch. A \textbf{71}, 1131 (2016).

\bibitem{Atteya2017} A. Atteya, \textit{et al.}, Eur. Phys. J. Plus \textbf{132}, 1 (2017).

\bibitem{C1} M.H. Rahman,\textit{et al.}, Phys. Plasmas \textbf{25}, 102118 (2018);
             N.A. Chowdhury, \textit{et al.}, Phys. plasmas \textbf{24}, 113701 (2017);
             M.H. Rahman, \textit{et al.}, Chin. J. Phys. \textbf{56}, 2061 (2018);
             N.A. Chowdhury, \textit{et al.}, Plasma Phys. Rep. \textbf{45}, 459 (2019);
             S. Jahan, \textit{et al.}, Plasma Phys. Rep. \textbf{46}, 90 (2020).
             
\bibitem{C2} S.K. Paul, \textit{et al.}, Pramana J. Phys. \textbf{94}, 58 (2020);
             N.A. Chowdhury, \textit{et al.}, Chaos \textbf{27}, 093105 (2017);
             N.A. Chowdhury, \textit{et al.}, Contrib. Plasma Phys. \textbf{58}, 870 (2018);
             M. Hassan, \textit{et al.}, Commun. Theor. Phys. \textbf{71}, 1017 (2019);
             D.M.S. Zaman, \textit{et al.}, High Temp. \textbf{58}, 789 (2020).

\bibitem{Mamun1999}  A. A. Mamun, Phys. Scr. \textbf{59}, 454 (1999).

\bibitem{Hossen2017b} M. M. Hossen, \textit{et al.}, High Energy Density Phys. \textbf{24}, 9 (2017).

\bibitem{Washimi1966} H. Washimi, T. Taniuti, Phys. Rev. Lett. \textbf{17}, 996 (1966).

\bibitem{C3} S. Jahan, \textit{et al.}, Commun. Theor. Phys. \textbf{71}, 327 (2019);
             N. Ahmed, \textit{et al.}, Chaos \textbf{28}, 123107 (2018);
             N.A. Chowdhury, \textit{et al.}, Vacuum \textbf{147}, 31 (2018);
             R.K. Shikha, \textit{et al.}, Eur. Phys. J. D \textbf{73}, 177 (2019);
             S. Banik, \textit{et al.}, Eur. Phys. J. D \textbf{75}, 43 (2021).

\end{thebibliography}
\end{document}